# Spatial and temporal dynamics of RhoA activities of single breast tumor cells in a 3D environment revealed by a machine learning-assisted FRET technique

**Brian CH Cheung[a], Louis Hodgson[b,c], Jeffrey E Segall[b] and Mingming Wu[a*]**

[a]Department of Biological and Environmental Engineering, Cornell University, Ithaca, NY, USA

[b]Department of Anatomy and Structural Biology, Albert Einstein College of Medicine, Bronx, NY, USA

[c]Gruss-Lipper Biophotonics Center, Albert Einstein College of Medicine, Bronx, NY, USA

[*]Corresponding author: mw272@cornell.edu

## ABSTRACT

One of the hallmarks of cancer cells is their exceptional ability to migrate within the extracellular matrix (ECM) for gaining access to the circulatory system, a critical step of cancer metastasis. RhoA, a small GTPase, is known to be a key molecular switch that toggles between actomyosin contractility and lamellipodial protrusion during cell migration. Current understanding of RhoA activity in cell migration has been largely derived from studies of cells plated on a two-dimensional (2D) substrate using a FRET biosensor. There has been increasing evidence that cells behave differently in a more physiologically relevant three-dimensional (3D) environment, however, studies of RhoA activities in 3D have been hindered by low signal-to-noise ratio in fluorescence imaging. In this paper, we present a machine learning-assisted FRET technique to follow the spatiotemporal dynamics of RhoA activities of single breast tumor cells (MDA-MB-231) migrating in a 3D as well as a 2D environment using a RhoA biosensor. We found that RhoA activity is more polarized along the long axis of the cell for single cells migrating on 2D fibronectin-coated glass versus those embedded in 3D collagen matrices. In particular, RhoA activities of cells in 2D exhibit a distinct front-to-back and back-to-front movement during migration in contrast to those in 3D. Finally, regardless of dimensionality, RhoA polarization is found to be correlated with cell shape.

## HIGHLIGHTS

- Robust machine learning approach for FRET image preparation and analysis
- Revealed spatial and temporal dynamics of RhoA activities of single cells in 2D and 3D
- Polarization strength and cell shape are correlated for cells in 2D and 3D

## INTRODUCTION

Cell migration within a three-dimensional (3D) extracellular matrix plays a critical role in many physiological processes (e.g. immune response and wound healing) as well as pathological processes (e.g. cancer metastasis and fibrosis)[1-2]. To metastasize, cancer cells need to move away from the primary site, into interstitial space and gain access to the circulatory system[3-4]. Traditionally, cell migration studies have been performed when cells are plated on a 2D substrate [5]. With the advancement of biomaterials and microfabrication technology, we now know that what we have learnt of cells in 2D does not always apply to those in 3D[6]. Broadly speaking, mammalian cell migration within a 3D ECM has been grouped into the following three categories: (i) amoeboid motility, where cells migrate in a path-seeking manner with diffuse, short lived, integrin-independent adhesion and then squeeze through the ECM pores[7]; (ii) an integrin dependent and matrix metalloproteinases (MMPs)-independent mesenchymal motility where cells adhere to pull on collagen fibers, and contract to move through the matrix when pore size is large enough for passage of the nucleus[7-8]; (iii) an integrin and MMPs-dependent path creating migration where cells use MMPs to digest the matrix in the front and migrate through the self-dug micro-channels using adhesion and contraction[6,9-13]. Central to all types of cell motility is the cell contractility.

To enable migration, RhoA GTPase, one of the canonical Rho-family p21 small GTPases, acts as a crucial molecular switch that controls downstream cytoskeletal effectors. Antagonized by another small GTPase, Rac1, RhoA triggers actin polymerization and actomyosin assembly through activation of mammalian Diaphanous (mDia)-related formins and Rho-associated protein kinases (ROCKs)[13]. RhoA and Rac1 cycle between GTP-bound (active) state and GDP-bound (inactive) state under the regulation of guanine nucleotide exchange factors (GEFs) and GTPase activating proteins (GAPs)[14]. RhoA GTPase is also known to regulate actomyosin contractility, cell morphology and migration through interactions with focal adhesion kinase (FAK)[15-17]. In lamellipodium-driven migration, while Rac1 predominantly acts to form protrusions, its counterpart RhoA is activated also at the leading edge and the lateral edges of lamellipodia which may be important for protrusion shaping[18-19]. On the opposite end of the cell, RhoA can be activated by RhoGEFs which leads to rear retraction[20]. RhoGTPases shuttle between the plasma membrane when activated and the cytosol in complex with the guanine nucleotide dissociation inhibitor (GDI) when inactivated[21]. Moreover, the localization and spatial distribution of RhoGTPases plays a key role during directed cell migration[22-27]. On one hand, local activation of RhoGTPases can drive directional cell migration, but, on the other hand, disruption of RhoGTPase localization can lead to general defects in motility[19,27-29].

In the context of cell migration, the spatial and temporal dynamics of RhoA and other RhoGTPases activity has been explored extensively[23,30-31]. The ratiometric Förster Resonance Energy Transfer (FRET) biosensors have played important roles in our ability to visualize the localization of the active and inactive forms of RhoGTPases within living cells[19,23,31-34]. Here, we use a genetically encoded FRET biosensor containing the FRET donor and acceptor fluorescent proteins in a single-chain molecular backbone[23,32], in contrast to the FRET biosensors based on two separate molecular components that come together as a function of activation[35]. This arrangement ensures an equimolar distribution of the FRET donor/acceptor moieties everywhere within the cell[32], which allows the approximation of the relative FRET efficiencies within a given cell by a ratiometric analysis of the FRET emission to the donor emission[36]. In addition to the FRET sensor itself, a critical step of the ratiometric FRET calculation is that the FRET/CFP image pairs are accurately aligned and segmented. Previously, these image pairs had been segmented manually, limiting the robustness of tracking fine cellular features like protrusions. Although automated approaches including Otsu's thresholding and k-means clustering have gained popularity in rapid image segmentation, they are often challenged by a continuous gradient of fluorescence intensities within a cell and near cell edges close to background. To circumvent this limitation, we used a neural network assisted algorithm (U-Net, University of Freiberg)[37] to trace the cell outlines of FRET/CFP image pairs. This technique enabled a robust way to obtain cell outlines precisely, and allowed us to visualize spatial and temporal dynamics of RhoA activities of cells in 3D for the first time.

**MATERIAL AND METHODS**

**Cell Line and RhoA Biosensor**

Triple-negative breast tumor cells (MDA-MB-231), were transduced with retrovirus harboring the tetOFF tTA (pQCXIN-tetOFF tetracycline-TransActivator, second generation; Clontech) and selected for stable genomic integration using G418 at 1mg/mL in the culture media, as previously described[38-39]. Following establishment of the stable population, cells were transduced with retrovirus harboring the synonymously modified single-chain RhoA FRET biosensor, based on RhoA FLARE.sc[23,40], in the pRetro-X-Puro backbone (Clontech). The transduced cells were selected for stable genomic integration using 10µg/mL puromycin dihydrochloride (A1113802, Gibco) in culture media, together with 2ug/mL doxycycline hyclate (D9891, Sigma) to repress the biosensor expression under normal culture[32]. Cells were maintained at 37℃ and 100% humidity in a 5% $CO_2$ incubator and were cultured in T75 flasks (10062-860, Corning) using Dulbecco's Modified Eagle's Medium (DMEM) (15-013-CV, Corning) with 4.5 g/L glucose, sodium pyruvate supplemented with 1:100 GlutaMAX (35050061, Gibco), 10% fetal bovine serum (S11150, Atlanta Biologicals), 1% penicillin/streptomycin (15140122, Gibco), and 1 mg/mL G418 (10131027, ThermoFisher). Culture medium was changed every 2 days to maintain a reasonable level of doxycycline[32,41]. Cells were cultured in phenol red-free complete DMEM (17-205-CVR, Corning) 1 day prior to the experiment to reduce the background autofluorescence from the media during imaging. Cells were passaged when they reached 70-90% confluency. Cells with 19-20 passages were used for the experiments.

For induction of biosensor expression, 48hrs prior to experiments, cells were trypsinized and doxycycline was removed to allow expression of the biosensor by replacing medium with doxycycline-free complete medium. Cells were then plated at a density of 100,000 cells/T25 flask and were incubated for 48h. The cells were checked for YFP signal before experiments to ensure the expression of biosensor.

**Device Preparation**

For both 2D and 3D experiments, 50mm uncoated glass bottom dishes with No. 0 glass (D50-14-0-U, Matsunami) were used. For 2D experiments, cover glasses were coated with 1µg/cm$^2$ fibronectin[42-44] (F0556, Sigma-Aldrich) by incubating for 1hr at room temperature according to manufacturer's protocol. Note that this coating density is beyond the theoretical saturation density which ensures a complete coverage of fibronectin on the substrate. Although a fibronectin study conducted by DiMilla et al. recommended an optimal coating density of 0.073µg/cm$^2$ for migration of human smooth muscle cells (HSMCs)[44], we found that, under 1µg/cm$^2$ fibronectin, the velocity of MDA-MB-231 in the presence of 20 ng/mL epidermal growth factor (EGF) was comparable to the peak velocity (~60µm hr$^{-1}$) demonstrated by DiMilla et al. Following incubation, fibronectin solution was aspirated and rinsed with media. The coated dish was used the same day after the coating process. For 3D experiments, polydimethylsiloxane (PDMS) microwells were prepared by mixing 10g silicone elastomer with 1g curing agent thoroughly. The mixture was degassed in vacuum for 20mins to remove air bubbles, followed by incubation at 60 ℃ overnight for polymerization. Three through holes with a diameter of 2 mm were created using a biopsy punch on the PDMS sheet with a thickness of 300 µm. PDMS sheet and cover glass were bonded following plasma treatment with an oxygen plasma oven (PDC-001, Harrick Plasma), followed by autoclave sterilization. To enhance attachment of collagen matrices to the glass/PDMS surfaces, cover glasses were treated with 5 µL of 1% polyethyleneimine (PEI) for 10 mins and then with 5 μL of 0.1% glutaraldehyde for 30mins. The dishes were left in biohood overnight in sterile distilled water and were washed with phosphate buffered saline (PBS) and aspirated before experiments.

**2D Cell Seeding**

Pre-coated glass bottom dishes were pre-warmed at 37℃ before seeding. Before experiments, cells were serum-starved overnight (~16h). Serum-starved cells were resuspended in serum-free medium, and plated at a density of ~4280 cells/cm$^2$. The plated cells were incubated at 37℃ and 5% $CO_2$ for 1hr for attachment. Medium was then replaced with fresh medium with 20 ng/mL epidermal growth factor (EGF) (AF-100-15, PeproTech) to enhance motility and cells were settled for 2hrs prior to imaging.

**3D Cell Culture**

Type I collagen extracted from rat tail tendon (354249, Corning) was suspended in 0.1% acetic acid (9.41 mg/mL). On ice, 31.9 μL collagen stock (9.41 mg/mL), 20ug fibronectin (F0556, Sigma-Aldrich), and 0.2 μL 1N NaOH were mixed with cell suspension in complete medium to reach a final volume of 200 μL. The final collagen concentration was 1.0 mg/mL and cell density was 150 cells/μL. 5 μL of the cell embedded collagen mixture was placed into the each of the three PDMS microwells (2mm diameter and 300 $\mu$m in depth) for polymerization in a 37℃ and 5% $CO_2$ incubator for collagen polymerization. To prevent cells from sinking to the bottom of the glass bottom dish during polymerization, the device was first placed upside-down for 5 mins. Subsequently, the dish was flipped twice at time points 15 min and 30 min. Note that we lifted the glass above any incubator surfaces to adopt a slow warming procedure for polymerization to obtain long and thick collagen fibers in contrast to the fast warming process described in our previous publication that yields thin and dense collagen fibers[45]. Throughout the process, the glass bottom dish was contained within a second moist petri dish to prevent collagen dryout. After polymerization, 1.5mL complete medium with 20 ng/mL EGF (AF-100-15, PeproTech) was added to each glass bottom dish. Cells were settled for 2hrs prior to imaging.

**FRET Image Acquisition**

We denote hereon, images from the donor-excitation-acceptor-emission as "FRET", and to avoid confusion, the final ratiometric FRET images are denoted as "rFRET". The optical setup is shown in Fig. S1A. The epifluorescence microscope was encased in an incubator with a temperature of 37℃, 5% $CO_2$, and humidity of ~70%. All CFP-FRET image pairs were taken with a 60x magnification water immersion objective lens (NA = 1.2; UPLSAPO60XW, Olympus America) installed on an epi-fluorescence microscope (IX81, Olympus America) and a 16-bit sCMOS camera (ORCA Flash 4.0 V3, Hamamatsu Photonics). Immersol W 2020 ($\eta$ = 1.3339; 444969, Carl Zeiss), a non-evaporative immersion medium, was used to match optical indices between the sample and objective lens. The light source for fluorescence imaging was a xenon arc lamp (Lambda LB-LS/30, Sutter Instruments). An absorptive neutral density filter with OD = 0.3 (NE03B-A, ThorLabs) was used to attenuate light to minimize phototoxicity. To obtain CFP and FRET signals simultaneously for ratiometric FRET calculation, the W-View Gemini beam splitter was used (A12801-01, Hamamatsu Photonics). The optical filters used were as follows: ET436/20x (CFP excitation), ET480/40m (CFP emission); ET500/20x (YFP excitation), ET535/30m (YFP emission); T455lp (dichroic mirror for reflecting excitation light and relaying emission from cells), T505lpxr (dichroic mirror for splitting CFP and FRET signals), T515lp (longpass filter for checking sensor expression by YFP signals). All filters and dichroic mirrors were purchased from Chroma Technology. Prior to each experiment, the $x$-$y$ positions of CFP and FRET channels were coarsely adjusted using the W-VIEW Adjustment software (Hamamatsu Photonics) to achieve a reasonable alignment for easier image registration and processing.

The glass bottom dish was placed on an x-y microscope stage (MS-2000, Applied Scientific Instrumentation), and images were taken every 7 min for 133 mins using Metamorph (version 7.7.7.0) (Molecular Devices) with an exposure time of 800ms. At each x-y position, a total of 20 images were taken. We constrained both the exposure time and experiment time to limit phototoxic effects on cells. To effectively maintain a consistent focal plane at this high magnification (60x) setting, between each image acquisition, an image-based automatic step focusing procedure in MetaMorph (Search type: Step; Algorithm: Standard; Range +/-: 1 µm; Max. step size: 0.2µm) was carried out with a minimal exposure time of 50ms to avoid photobleaching. In both 2D and 3D setting, cells were tracked in the focal plane and the measured speed is the actual speed of cells in the focal plane.

**Field Alignment and Correction**

Field alignment and flatfield correction were performed before FRET imaging acquisition.

*Field alignment*

To align the CFP/FRET image channels from misalignments caused by the difference in optical paths (Fig. S2A,B), we conducted the following calibration procedures. First, a target pattern provided by Olympus America was imaged under brightfield as a guideline for rough cropping (Fig. S1B). The center of the target served as a reference for aligning the two channels. To further align and map the channels, a glass slide coated with 4-$\mu$m fluorescence marker beads of 4 different colors, blue, green, orange, and dark red (T14792, Invitrogen) was imaged in fluorescence mode to produce reference images from the CFP and FRET channels. Then, the bead image was cropped and split into two images according to the reference point calculated earlier (Fig. S1B). The split bead images were then used to generate registration parameters using the affine transformation algorithm from MatLab to achieve subpixel accuracy in field alignment. This feature-based algorithm corrects for $x$-$y$-translation, rotation, scaling, and shear. We were able to reduce misalignment down to less than 1 pixel (Fig. S2C).

*Flatfield correction and background subtraction*

Flatfield correction and dark current (DC) subtraction described by Spiering et al[32] were performed after field alignment. For dark current subtraction, 10 DC images were taken from the field of view with no illumination and a closed shutter. To compensate for spatial unevenness in illumination of the field of view, 10 shade images of a dish of medium with the same optical setup as the experiments were taken. Both DC images and shade images were averaged respectively for calibration of the two channels. The following formula was used to calculate the corrected image[32]:

$$[image]_{corr} = \frac{[image]_{raw} - [DC] + \overline{[DC]}}{[image]_{shade} - [DC] + \overline{[DC]}} \times Scaling\ factor$$

Since we were taking a ratiometric calculation, the choice of scaling factor did not affect our final calculation as long as it was consistent for both channels. We used a scaling factor of 3000 to bring images to visualization under a 16-bit domain.

After flatfield correction, we selected a blank area away from the cell and any cell debris to measure its average intensity value. This was done for each individual frame as the background intensity can change in a time series movie. The average intensity of the blank area was then subtracted from the flatfield corrected image.

**Cell Segmentation and Tracking**

A convolutional neural network package, U-Net, developed in the University of Freiburg[37] was adopted to segment the epi-fluorescence images of the cells. Details of this segmentation method is illustrated in Results and Discussion section. To track cell movements, the centroids of the binary masks of the segmented cells were defined as the center of cells.

**Ratiometric FRET Calculation and RhoA biosensor validation**

The FRET channel was divided by the CFP channel, then multiplied by a scaling factor to bring the image back to the 16-bit dynamic range for visualization purposes. The ratiometric FRET image, rFRET, was calculated using the following equation:

$$[rFRET] = \frac{[FRET]_{corr}}{[CFP]_{corr}} \times Scaling\ factor$$

where $[FRET]_{corr}$ and $[CFP]_{corr}$ are the corrected images of FRET and CFP channels respectively.

The final images were then corrected for photobleaching using the biexponential intensity decay model described in Spiering et al[32,34].

To validate the cell line with the FRET biosensor, we modulated RhoA activity by treating cells with CN03, a constitutive RhoA activator (CN03, Cytoskeleton, Inc.)[46], and Rhosin (Fig. S3), a RhoA inhibitor[47]. CN03 activates RhoA GTPase by deamidating glutamine-63, which is located at the Switch II region[46]. This allows constitutive activation of RhoA without altering the availability of Switch I region for binding with Rho-Binding Domain (RBD) within the biosensor. On the other hand, Rhosin (555460-M, Millipore) is an RhoA inhibitor which targets the RhoGEF binding domain of RhoA[47]. We found that Rhosin treatment reduced the average FRET ratio in MDA-MB-231, while cells treated with CN03 had a higher average FRET ratio compared to those treated with Rhosin (Fig. S3).

**Measurement of Polarization of RhoA Activities and Motility Characterization**

We used the first spatial moment of rFRET signal, $\vec{P}$, to represent the polarization of RhoA activities, analogous to the definition of electric dipole moment in physics[48], which is commonly used to quantify the polarization of an electric charge distribution. As shown below, RhoA polarization $\vec{P}$, is the sum of the product of rFRET intensity at a given point and the displacement with respect to the centroid of the cell (Fig. S4A):

$$\vec{P} = \sum_{i=1}^{n} I_i \cdot (\vec{r_i} - \vec{r_c})$$

where $I_i$ is the rFRET intensity at position $\vec{r_i}$, $\vec{r_c}$ is position vector of the centroid of the cell extrapolated from elliptic fit using ImageJ. $n$ is the total number of pixels within the cell. Polarization strength is represented by $|\vec{P}|$.

After elliptic fit was performed, component vectors of velocity $\vec{V}$ and polarization $\vec{P}$ were calculated with reference to the cell's long (L)-axis and short (S)-axis (Fig. S4B). Here, the angle between $\vec{V}$ and $\vec{P}$ is defined as the circular distance between two vectors in counterclockwise direction. Finally, to distinguish between amoeboid and mesenchymal motility, a cell is considered mesenchymal when its aspect ratio (i.e. $\frac{L_x}{L_y}$) is greater than or equal to two, to be consistent with previous work[49-50].

**Data Analysis**

Aspect ratios of cells were generated from elliptical fit using ImageJ (NIH). Polarization of RhoA activities, cell speed, and cell trajectories were computed using in-house MatLab programs. All statistical analyses were performed using Prism 8 (GraphPad) and in-house Matlab codes. Student's t tests were performed on data sets which had similar variances, and Welch's t tests were performed on data sets with unequal variances. The F test was used to compare variances between data sets. Spearman correlation analyses were performed in all correlation studies. To test for significance between correlation studies, we compared the coefficient $\rho$ with null hypotheses that $\rho_0 = 0$. For analyses between velocity or speed versus polarization, we used a polarization value one time-point before velocity or speed, i.e. $P$ at t = 0min was paired with $V$ at t = 7min.

For angular data, the Rao's spacing test and resultant vector length were used to study the uniformity of vector orientations[51]. To define the resultant vector length in directional data analysis, we assign unit vectors $\widehat{x_1}, \cdots, \widehat{x_n}$ to corresponding angles $\theta_i, i = 1, \cdots, n$. The mean direction $\bar{\theta}$ of $\theta_1, \cdots, \theta_n$ is the direction of the resultant vector from $\widehat{x_1} + \cdots + \widehat{x_n}$, which is also the direction of the center of mass $\bar{x}$ of $\widehat{x_1}, \cdots, \widehat{x_n}$. The Cartesian coordinates of the center of mass $\bar{x}(C, S)$ are therefore $\left(\frac{1}{n}\sum_{k=1}^{n} cos\theta_k, \frac{1}{n}\sum_{k=1}^{n} sin\theta_k\right)$. The resultant vector length is given by $R = \sqrt{C^2 + S^2}$.

**RESULTS AND DISCUSSION**

**A machine learning-assisted cell image segmentation method for FRET signal computation**

Ratiometric computation of the FRET signal involves dividing the fluorescence cell image of the FRET channel by that of the CFP channel (Fig. 1). An important component of this computation is the accurate segmentation of cell images with weak fluorescence signal. We adopted a machine learning cell segmentation method previously developed for brightfield cell images to accurately identify the outlines of the cells in CFP-FRET image pairs[37]. Fig. 1 illustrates the procedures that we used to segment an image pair. (i) 10 images from the FRET channel (after alignment and correction) with manually drawn outlines together with a pre-trained model provided by the U-Net package were used as input to the U-Net finetuning module, and the outcome was an adapted model (Fig. 1A). (ii) An original FRET cell image along with the adapted model was used as the input to the U-Net segmentation model, and the output was the binary mask of the cell (Fig. 1B). (iii) Finally, the ratiometric FRET image was generated by taking a pixel-wise division of the intensity matrices of FRET and CFP images, and multiplied by the binary mask (Fig. 1C). We note that FRET images, in contrast to the CFP images, were used for model training and segmentation because of their higher signal-to-noise ratios. This process has been used successfully to automatically analyze live cells from time series of images as well as slices of images for 3D reconstruction.

Central to FRET image analysis is accurate segmentation as well as cell image alignment. Traditionally, segmentation of a cell image is done manually, which leads to person-to-person variability and misinterpretation of fine features such as cell protrusions which are critical for cell migration studies. For example, with a generally weak fluorescence signal and a histogram with no apparent bimodality, the Otsu method cannot accurately search for an optimal threshold that minimizes intra-class variance[52]. Despite requiring less computational power, the other commonly used segmentation method, k-means clustering, has also been challenged by low contrast situations where centroids of intensity clusters become proximal to each other[53]. The machine learning assisted FRET segmentation method has enabled us to successfully identify the outlines of cells in an automatic and consistent way. A second important feature of our FRET analysis is the alignment of cell images. The two images (CFP and FRET) are not always perfectly superimposable due to physical factors such as optical path differences introduced by the beam splitter or slight mechanical misalignment between filter cubes and curvature differences between dichroic mirrors (Fig. S2A). This misalignment could cause extreme ratios at the cell edge and miscalculated ratiometric FRET signal (Fig. S2B). Therefore, it is important to align images in a reproducible way. Here, we used multi-fluorescent beads to generate reference images in CFP and FRET channels and aligned image fields by affine transformation, which has enabled us to achieve image alignment at a subpixel accuracy (Fig. S2C).

**RhoA activities are more polarized in cells plated on 2D surfaces than those embedded within a 3D collagen matrix**

To follow the spatial and temporal dynamics of RhoA activities of malignant breast tumor cells (MDA-MB-231), we took time series of CFP-FRET image pairs of MDA-MB-231 cells with the RhoA biosensor in both 2D and 3D environments. RhoA activities were more polarized in 2D than those in 3D as seen in the ratiometric FRET images (Fig. 2A, Supplementary Video 1). In addition, RhoA activities of cells had a clear oscillation movement in a front-to-back and back-to-front fashion in 2D in contrast to that of in 3D (Supplementary Video 1-2).

To gain a quantitative understanding on how RhoA polarization is correlated with cell migration, we computed the polarization of RhoA activity, $\vec{P}$, using the rFRET images shown in Fig. 2A. We define RhoA polarization, $\vec{P}$, to be the first moment of the rFRET signals with respect to the centroid of the cell (Fig. S4). $P_L$ and $P_S$ are the component of $\vec{P}$ along the long (L)- and short (S)-axis of the cell. Fig. 2B illustrates an example of the time evolution of $P_L$ of a cell in 2D versus 3D. It shows that the cell in 2D had a larger $P_L$ oscillation magnitude than that in 3D. This observation is consistent with the computation of $P$ for multiple cells in 2D versus 3D (Fig. 2C-D) where the oscillation magnitude of RhoA activity for cells in 2D was significantly higher than that in 3D (Fig. 2C-E). Fourier transform of $|P|$ in 2D and 3D are consistent with Fig. 2C-D, in that a few cells in 2D showed clear periodic movement with a period of ~30 minutes while no discernible peak was observed from the cells in 3D (Fig. S6A and B).

It is interesting to note that, in 2D, cells were mostly elongated and adopted mesenchymal motility; while in 3D, cells were more rounded and exhibited an amoeboid motility (Fig. 2A). With a mesenchymal phenotype, the RhoA activities of the cell in 2D exhibited an apparent oscillating behavior where cell extension and lamellipodia formation was locally associated with weak RhoA activity, and cell contraction appeared to be accompanied by strong local RhoA activity (Fig. 2A). This behavior is less evident in cells in 3D (Fig. 2A). While Costigliola et al suggested that RhoA regulates calcium-independent periodic contractions of the cell cortex[54], it will be interesting to explore whether this behavior is applicable to all mesenchymal and amoeboid motility in the future.

**Shape and RhoA polarization of cells in 2D and 3D**

We calculated the cell aspect ratio (long axis/short axis) using cell images from the FRET channel. Fig. 3A and B show that cells in 2D had larger aspect ratio and polarization strength than those in 3D. To answer the question whether elongated cells are more polarized, we computed the Spearman correlation coefficient between $|P|$ and aspect ratio (Fig. 3C,D). Here, $\rho = 0$ means no association, $\rho = +1$ indicates a perfect positive monotonic relationship, and $\rho = -1$ indicates a perfect negative monotonic relationship. The results show that, in both 2D and 3D, cells with high RhoA polarization tend to be more elongated (Fig. 3C and D). This observation is consistent with Zemel et al, in which elongated cells tend to exhibit aligned stress fibers which are largely regulated by RhoA during actin polymerization[55-56]. To answer the question whether RhoA polarization follows the long axis of the cell, we computed the angle, $|\theta_{P_L}|$, between polarization vector $\vec{P}$ and the long axis (Fig. S7). A plot of $\cos(|\theta_{P_L}|)$ shows that polarization vectors in 2D were significantly more aligned with the long axes, compared to 3D (Fig. S7).

In summary, we see that RhoA activities were more polarized and cells were more elongated in 2D than 3D. This differential behavior may be explained by the distinct cell-environment interactions for cells in 2D versus 3D. In 2D, cells are plated on a rigid glass substrate where cells adhere to the fibronectin coated substrate via focal adhesions[57-58]. This allows greater cellular traction force which orients and aligns stress fibers, and subsequently leads to a polarized and an elongated phenotype[58-59]. In 3D, cells experience less adhesion to the matrix which means less traction force and less stress fiber alignment. This leads to a less polarized and more rounded cell morphology. Nonetheless, we note that the nature of the 3D matrix and cell type can impact cell shape and mode of migration[59-60]. For instance, migration among elongated cells can be driven by small lateral blebs called lobopodia which, in consistent with other contractility-based motility, are governed by RhoA, ROCK, and myosin II[59,61].

**RhoA polarization occurs at the rear-end of the cell during migration in 2D but not evident in 3D**

To delineate the relationship between migration direction and RhoA polarization of cells in 2D and 3D, we measured $\theta_{VP}$, the angle between velocity vector $\vec{V}$ and polarization vector $\vec{P}$ as defined in Fig. S4B. The polar plot for the distribution of $\theta_{VP}$ peaks at 180º for cells in 2D, which means cells had a tendency to move towards the direction that was opposite to its RhoA polarization. Meanwhile, the distribution of $\theta_{VP}$ was uniform in all directions for cells in 3D. We also computed the resultant vector length to evaluate whether $\vec{P}$ had a preferred direction relative to $\vec{V}$. Here, a resultant vector length of 0 would indicate non-directionality, and a length of 1 represents a strong preference of direction. As revealed by the resultant vector length and Rao's spacing test, polarization vectors $\vec{P}$ are oriented away from the direction of cell migration in 2D, but not in 3D (Fig. 4A and B).

These findings suggest that RhoA is mostly activated at the rear-end of the cell in 2D in a directional manner, but not in 3D. To explain, as cells are more capable of forming stable adhesions and protrusions with a rigid substrate in 2D, it is conceivable that polarization and actin alignment are more favorable for cells in 2D than in 3D. In breast cancer, short filopodium-like protrusions (FLPs), controlled upstream by RhoA-ROCK signaling, have been shown to be more abundant in the metastasis-competent MDA-MB-231 cell line[62]. While these RhoA-ROCK-driven short FLPs are responsible for initiating migration direction[63], we speculate that, in 3D, as cell adhesions are less stable due to a soft fibrous matrix, the regulation of RhoA activity via adhesion molecules such as integrins becomes less stable, which may result in less actin alignment. We also suspect that the requirement for RhoA polarization in 3D may indeed vary with matrix fiber density, as Lammermann et al reported that myosin II-inhibited dentritic cells can still migrate and drag the rear cell body in low density-collagen gels[9], even when contraction is blocked. And therefore, it will be interesting to study how tumor cells dynamically orchestrate their RhoA activity at different fiber densities as they often migrate in highly heterogenous metastatic niches.

**Cells migrate faster in 2D than 3D, and cell speed and RhoA polarization strength are poorly correlated**

We measured the speed of cells in 2D versus 3D, and investigated its correlation with RhoA polarization strengths. Fig. 5A shows maximum speed of cells in 2D versus 3D. In agreement with our previous work[64], cells in 3D migrated at a significantly lower speed than in 2D (Fig. 5A), and the speed was consistent with previous work using collagen matrices with similar concentrations (1.0mg/mL)[64]. The difference in speed between 2D and 3D can be explained by that, in a collagen matrix, cells have to overcome steric resistance from fibers and undergo nuclear deformation before they can migrate to a new position, whereas cells can form stable focal adhesions to exert force on the planar substrate and migrate in a 2D environment[65-67].

To test whether migration speed is associated with RhoA polarization, we performed Spearman coefficient analysis between speed and polarization. Interestingly, weak (3D) or no correlation (2D) were found between these two parameters (Fig. 5B, C). We conjecture that, polarization strength becomes less important for cell migration in 2D as cells use integrin dependent adhesion force to push forward on a stiff and stable substrate, in contrast to a soft and dynamic 3D fibrous network. Together with low aspect ratios, these findings suggest that cells in 3D may have adopted the "amoeboid" motility which often features high RhoA requirement[61,68].

**CONCLUSIONS**

In this study, we utilized a machine learning-assisted FRET analytical technique for studies of RhoA dynamics in both 2D and 3D platforms. This technique can potentially facilitate consistent and efficient analytical work on a high volume of FRET images, in particular, 3D reconstruction of cells embedded in collagen matrices to gain biophysical insights on how RhoGTPases like RhoA are modulated and generate contractile forces on the matrix during migration. In our experiments, we found that MDA-MB-231 cells were mostly elongated in a 2D glass substrate while cells in a 3D collagen matrix had a rounded morphology. During cell migration, compared to 3D, stronger oscillations of RhoA polarization activities were observed in 2D. We showed that while polarization is associated with cell elongation in both 2D and 3D, RhoA activities tend to align better with the long axis of cells in 2D, compared to 3D. Finally, cell tracking results show that cells moved faster in 2D than in 3D. While RhoA polarization is found to be weakly correlated with cell speed in 3D and not in 2D, further molecular studies are needed to delineate the mechanism by which RhoA polarization differentially affects cell migration speed and direction in 2D and 3D.

**ACKNOWLEDGEMENT**

This work was supported by grants from the National Cancer Institute [Grant No. R01CA221346 (MW and JES)], National Institute of General Medical Sciences [Grant No. R35GM136226 (LH)]; and by the Cornell Center on the Microenvironment & Metastasis [Award No. U54CA143876 from the National Cancer Institute]; the Cornell NanoScale Science and Technology, and the Cornell BRC imaging facility. JES is the Betty and Sheldon Feinberg Senior Faculty Scholar in Cancer Research. LH is a Irma T. Hirschl Career Scientist. We thank Dr. Yu Ling Huang and Young Joon Suh for their assistance in experiments. We also thank Dr. Thorsten Falk from the University of Freiburg for advice on the U-net software package. All Matlab codes and MetaMorph routines are available upon request.

**ABBREVIATIONS**

| | | |
|---|---|---|
| **2D** | (Two-dimensional) |
| **3D** | (Three-dimensional) |
| **CFP** | (Cyan fluorescent protein) |
| **DC** | (Dark current) |
| **ECM** | (Extracellular matrix) |
| **EGF** | (Epidermal growth factor) |
| **FAK** | (Focal adhesion kinase) |
| **FRET** | (Förster resonance energy transfer) |
| **GAP** | (GTPase activation protein) |
| **GDI** | (Rho GDP-dissociation inhibitor) |
| **GEF** | (Guanine nucleotide exchange factors) |
| **mDia** | (Mammalian Diaphanous-related Formin) |
| **MMP** | (Matrix metalloproteinase) |
| **PBS** | (Phosphate buffered saline) |
| **PDMS** | (Polydimethylsiloxane) |
| **PIP3** | (Phosphatidylinositol (3,4,5)-trisphosphate) |
| **RBD** | (Rho-binding domain) |
| **RhoA** | (Ras homolog gene family, member A) |
| **RhoGTPase** | (Rho family guanosine triphosphatase) |
| **ROCK** | (Rho-associated Protein Kinase) |
| **YFP** | (Yellow fluorescent protein) |

**FIGURES**

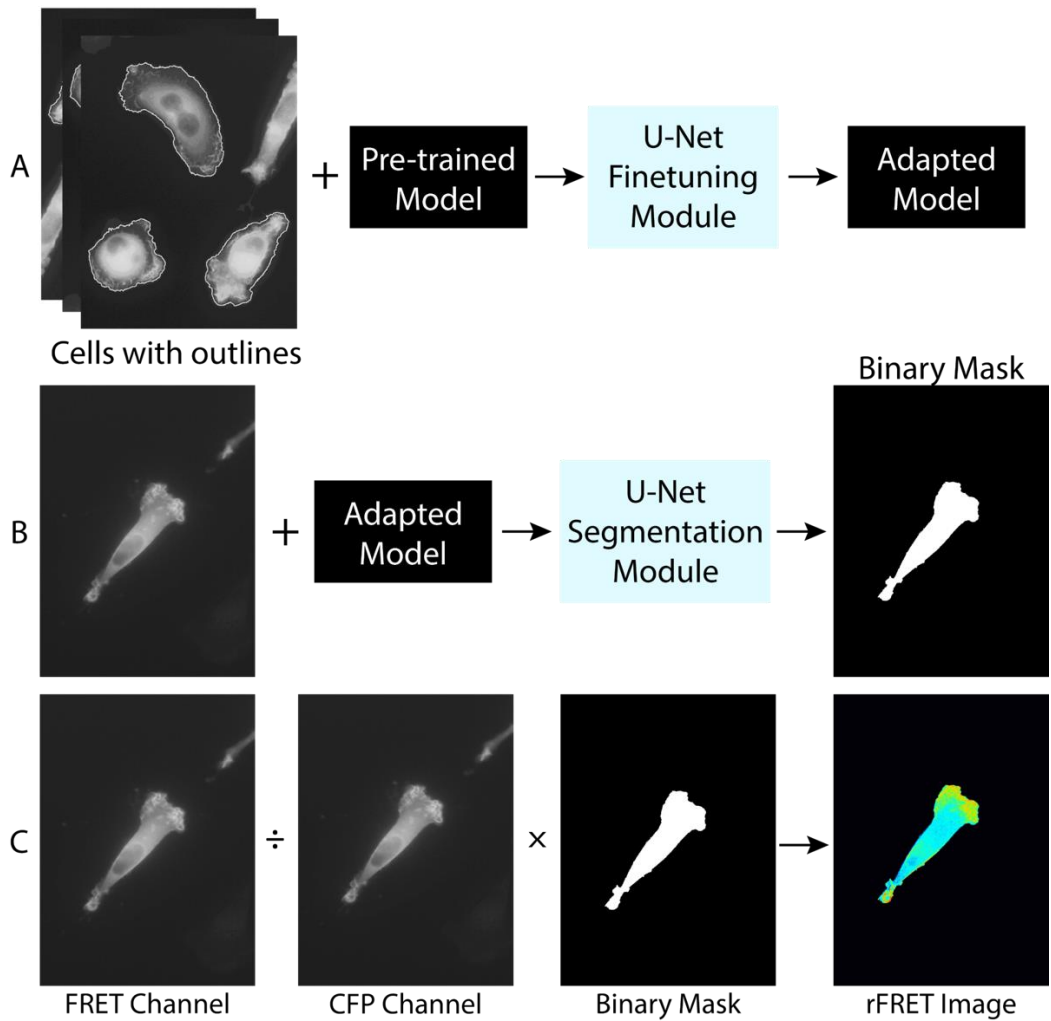

**Figure 1. Illustration of a machine learning enhanced FRET technique. A.** Model training using a convolutional neural network (U-Net) package. Original images of cells with manually drawn outlines together with pre-trained model are input to the U-Net finetuning module. The output is the trained adapted model. **B.** Cell binary mask generation using the adapted model. The cell image from FRET channel and the adapted model are fed into the U-Net segmentation module. The output is a binary mask of the cell. Here, cell images from the FRET channel are used. **C.** Ratiometric FRET calculation. The ratiometric FRET image is obtained by dividing the image of the FRET channel from that of the CFP channel, and then segmented by the binary mask. The final image is then corrected for photobleaching.

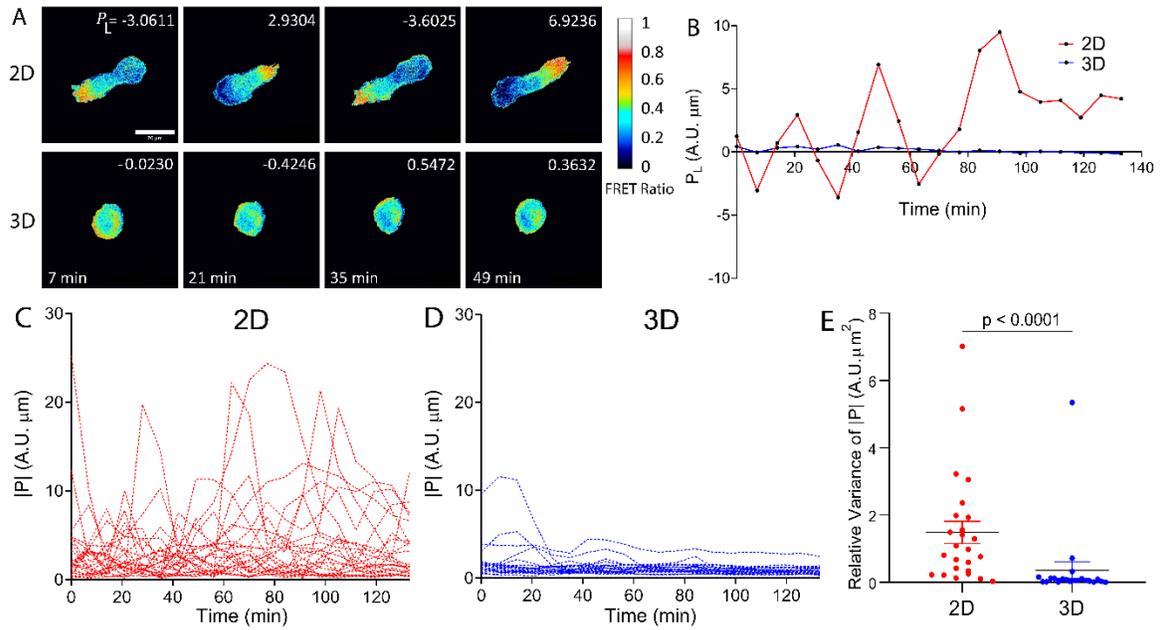

**Figure 2. Spatial and temporal dynamics of RhoA activities of MDA-MB-231 Cells in 2D versus 3D. A.** RhoA activities represented by the ratiometric FRET signal of the cells seeded in 2D versus in 3D. The color indicates the FRET ratio. Cells in 2D showed strong RhoA polarization activities and elongated morphologies, whereas cells in 3D remained rounded and polarization was not as pronounced. The legend in each image is the polarization magnitude along the long axis of the cell, $P_L$. The scale bar represents 20 $\mu$m. Images were taken at every 7 mins, and t=0 is defined as the starting point of imaging, which is about 2h after EGF treatment for motility enhancement. **B.** Computed polarization along the long axis of the cell, $P_L$, demonstrates a more pronounced oscillating movement in a front-to-back and back-to-front fashion during migration in 2D, in contrast to that of 3D. **C-D.** Temporal dynamics of RhoA activities of cells in 2D and 3D. Each line is time evolution of the absolute value of the polarization from one cell. **E.** While RhoA polarization oscillated in both 2D and 3D, the relative variance of polarization strength in 2D was significantly higher than that of 3D (F test, p < 0.0001).

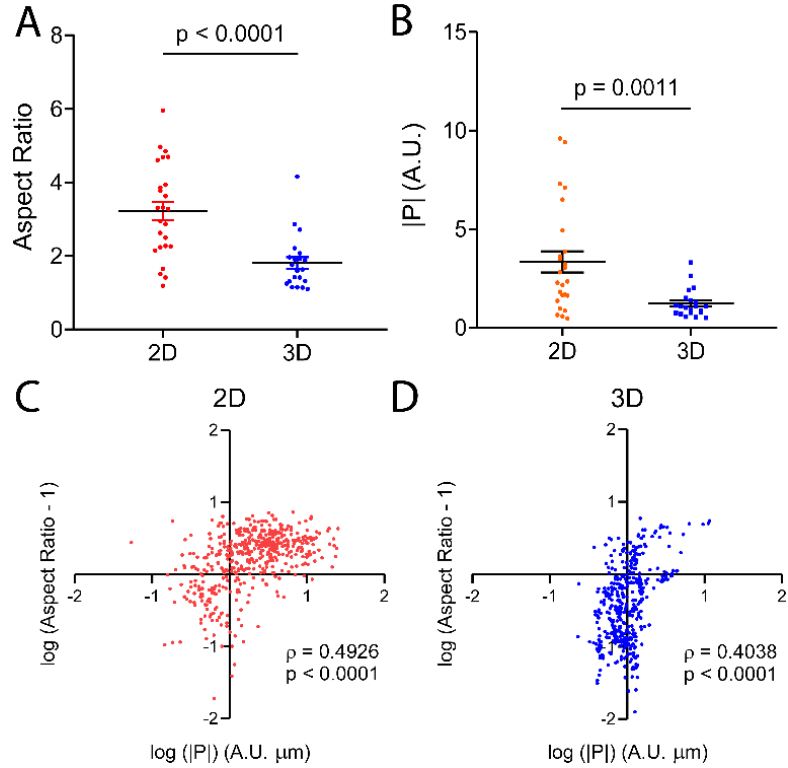

**Figure 3. Cell shape and the polarization of RhoA activities of the cell in 2D versus 3D. A.** Aspect ratio of individual cells in 2D versus 3D. Cells are more elongated in 2D than 3D. **B.** Magnitude of $P_L$ of cells in 2D versus 3D. Cells in 2D are significantly more polarized than those in 3D. Each dot represents the polarization strength $|P|$ of a cell. Error bar = SEM in both A, B. **C, D.** Scatter plots of log of (aspect ratio-1) versus log of polarization strength $|P|$ for cells in 2D (C) and 3D (D). Spearman correlation coefficient, $\rho$, between $|P|$ and aspect ratio at all instances during cell migration was computed. Here, $r = 0$ means no correlation, $r = +1$ indicates a perfect positive monotonic relationship, and $r = -1$ indicates a perfect negative monotonic relationship. Each dot represents a time point of a cell, with n = 500 for 2D, and n = 420 for 3D. 25 cells were followed for 2D, and 21 cells were followed for 3D. $\rho$ is the Spearman correlation coefficient. The *p*-value represents the significance of a statistical test of the computed correlation coefficient $\rho$ against the null hypothesis $H_0$ in which $\rho_0 = 0$. Cells in both 2D and 3D exhibited a positive relationship between polarization strength and cell elongation.

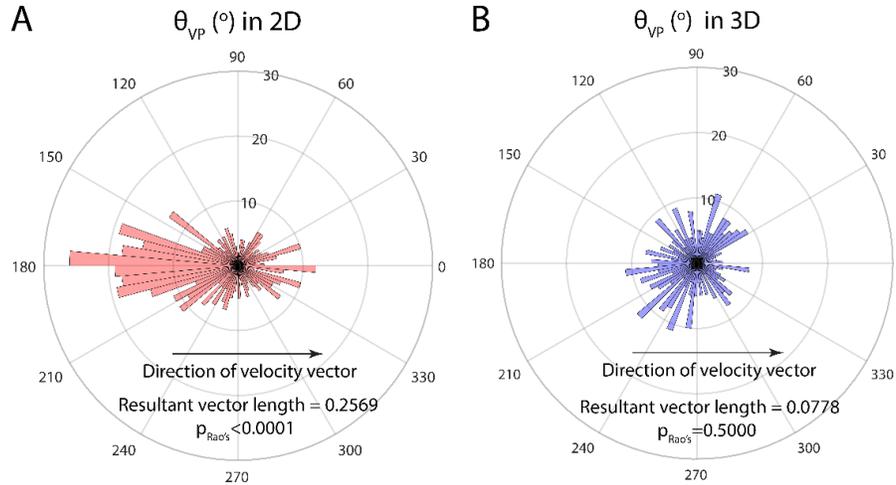

**Figure 4A-B. Directionality of RhoA polarization revealed by the Rao's spacing test between polarization and velocity.** Polar plot of the distribution of $\theta_{V_P}$, the angle between velocity and polarization vector in 2D (A) and 3D (B). In 2D, RhoA polarization peaks around 180° with respect to the direction of velocity (p < 0.0001, Rao's spacing test), indicating that cells tend to move against the direction of polarization. RhoA polarization in 3D exhibits no preferred direction (p = 0.5000, Rao's spacing test). Axis in radial direction shows the number of data points in a given direction (n = 475 for 2D, and n = 399 for 3D).

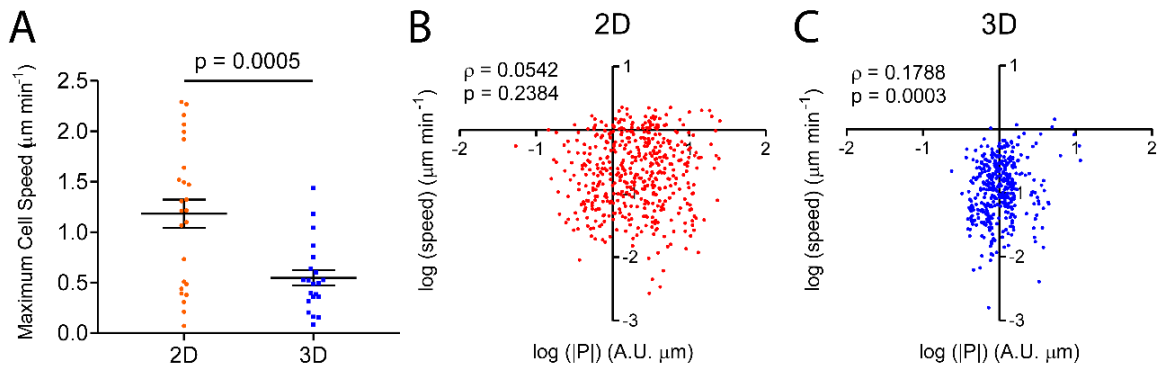

**Figure 5. Cells move faster in 2D than in 3D, cell speed and polarization strength are poorly correlated in 2D and 3D. A.** Cells in 2D migrate at a significantly higher speed than those in 3D. Error bar: +/-SEM. **B, C.** Scatter plots of speed versus polarization magnitude |P| for cells in 2D (C) and 3D (D). Speed and polarization strength are very weakly correlated (p<0.0001) in 3D, but not in 2D. $\rho$ is the Spearman correlation coefficient. The $p$-value represents the significance of a statistical test of the computed correlation coefficient $\rho$ against the null hypothesis $H_0$ in which $\rho_0 = 0$. n = 475 for 2D and n = 399 for 3D.

SUPPLEMENTARY MATERIALS

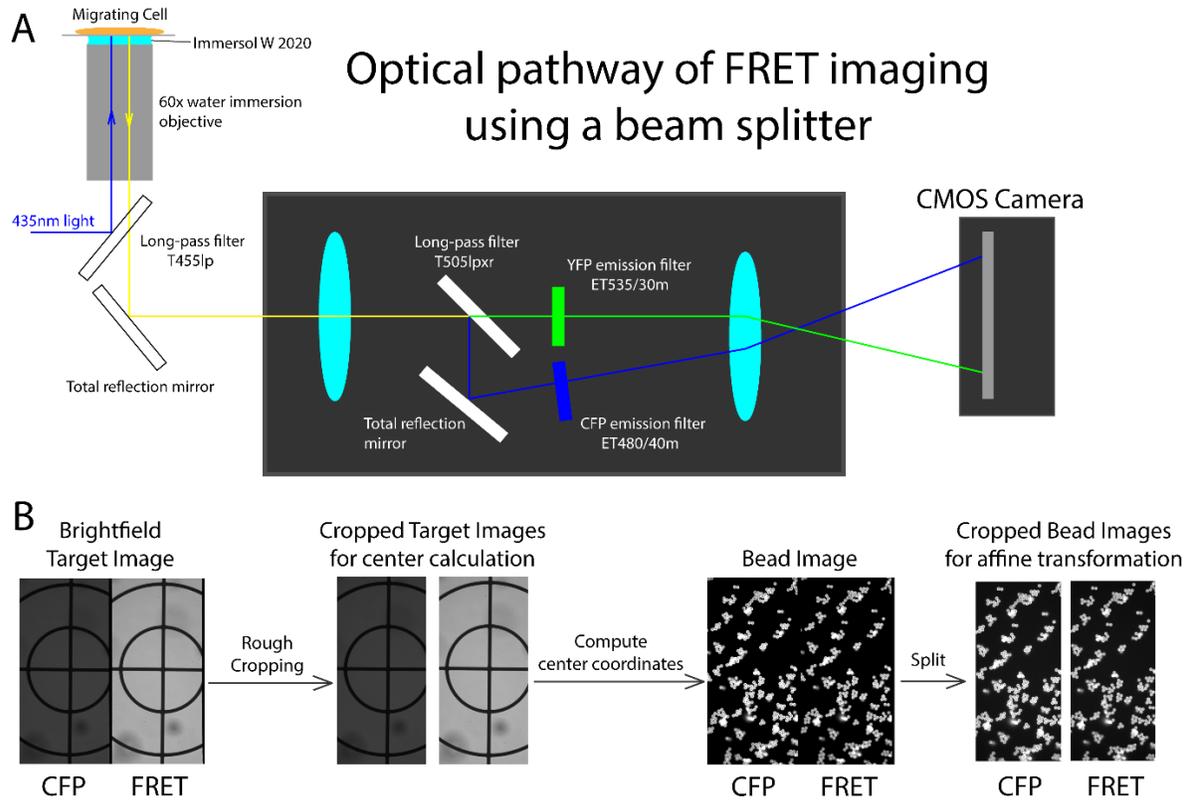

**Figure S1. FRET imaging workflow. A.** Optical path layout of the beam splitter for FRET imaging. **B.** Workflow chart for field alignment using a brightfield target image and then fluorescence bead image. A rough cropping is performed for computation of field centers of two channels using the brightfield target image. The calculated pixel coordinates are then applied on the bead image to generate two separate images for field mapping using affine transformation.

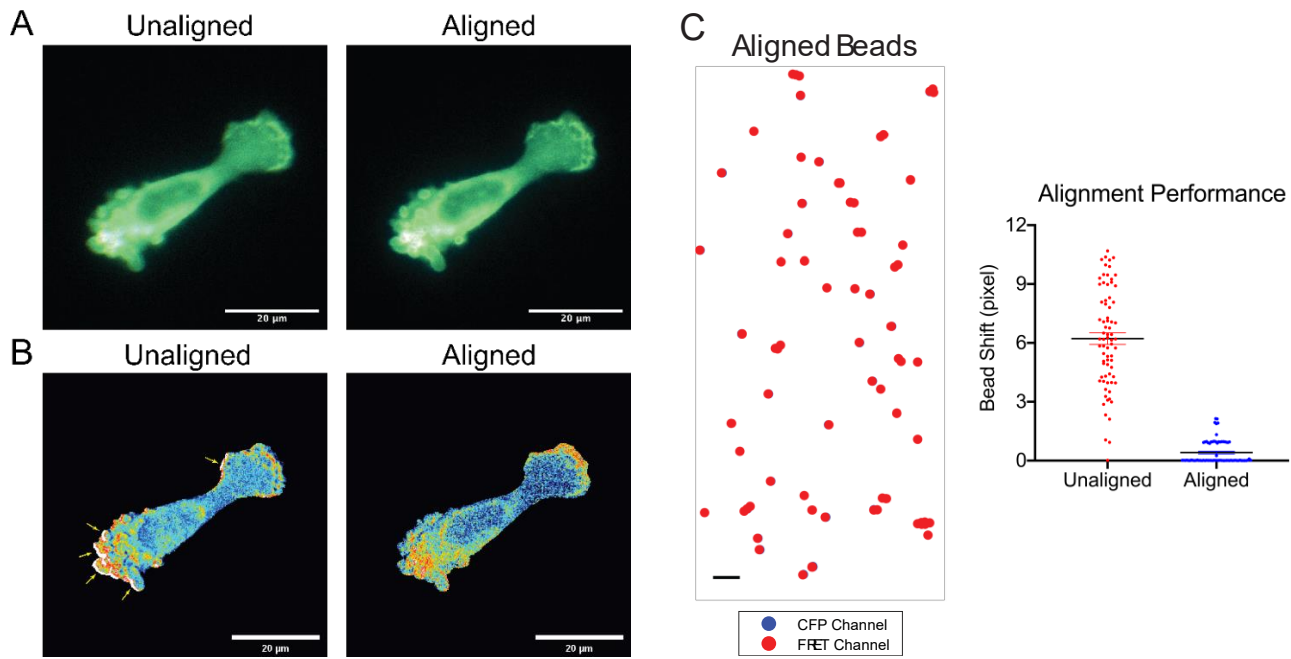

**Figure S2. Examples of aligned cell and bead images. A.** Overlayed cell images from CFP and FRET channel before (left) and after (right) alignment. Cell images were registered using the displacement field generated from affine transformation in MatLab. The displacement field was then applied onto individual pixels to align the images. Scale bar = 20$\mu$m. **B.** Ratiometric FRET images generated from before (left) and after (right) aligned CFP and FRET channel. Extreme FRET ratio due to misalignment causes edge artifacts (arrows). Scale bar = 20$\mu$m. **C.** Centroid plot of beads used in image alignment (left) and quantification of pixel shifts of individual beads (right). After image registration, the mean displacement of beads between channels reduced from 6.218pix to 0.412pix, which translates to a 0.045$\mu$m shift. Scale bar = 10$\mu$m. Error bar = SEM.

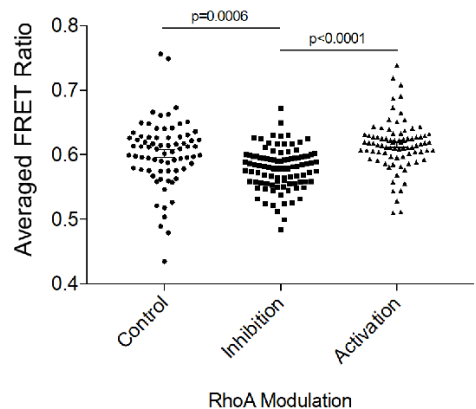

**Figure S3. FRET sensor was validated in 2D by activation and inhibition assays.** When treated with Rhosin, which targets the Trp58 residue situated at the LARG (one of the RhoGEFs) binding site of RhoA, FRET ratios of cells decreased significantly when compared to the control group. When treated with CN03, which constitutively activates RhoA by deamidating glutamine-63 located in the Switch II region while maintaining the availability of Switch I region of RhoA for RBD binding, FRET ratios of cells were significantly higher than inhibition group. Error bar = SEM.

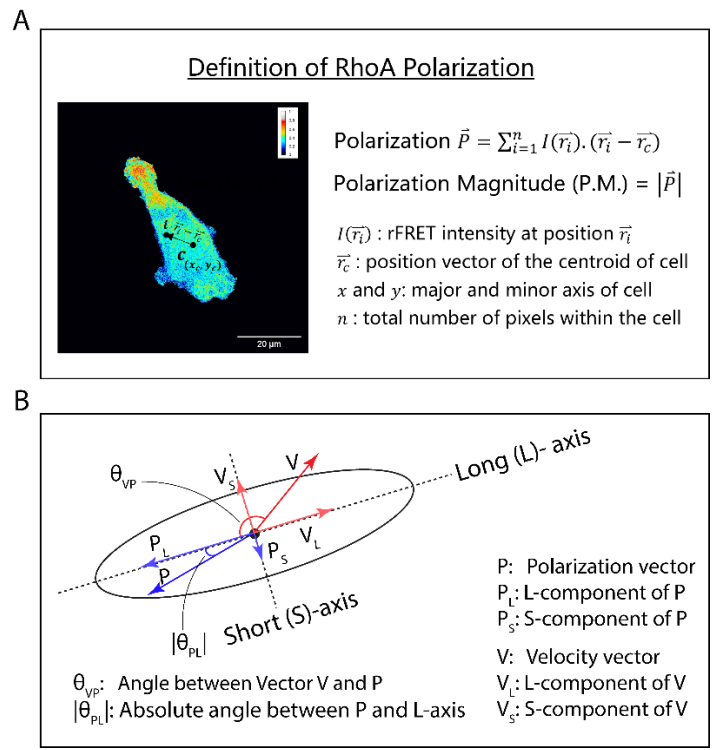

**Figure S4. Measurement and interpretation of polarization of RhoA activities. A.** RhoA polarization is defined by the sum of all products of displacement vector with respect to the centroid of the cell and intensity of individual pixels. Scale bar = 20$\mu$m. **B.** Long (L)- and short (S)- axes were generated from elliptic fit function using ImageJ. Velocity $\vec{V}$ and polarization $\vec{P}$ were decomposed along the cell's long (L)-axis and short (S)-axis. Here, the angle between $\vec{V}$ and $\vec{P}$ is defined as the circular distance between two vectors in counterclockwise direction. The deviation between $\vec{P}$ and L-axis is defined as the absolute angle between $\vec{P}$ and L-axis, with a range of 0 to 90 degrees.

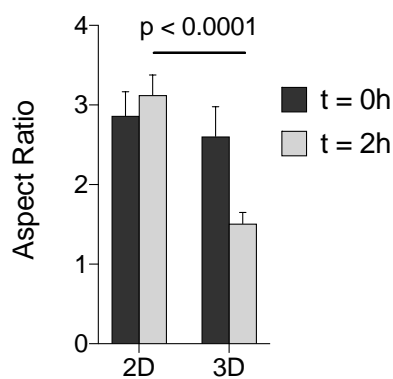

**Figure S5. Effect of EGF on aspect ratio.** To stimulate cell motility, we pre-treated cells with EGF 2 hours prior to the experiment. Cells embedded in a 3D environment had lower aspect ratios at the start of experiment, which is 2hrs after EGF treatment. Error bar = SEM.

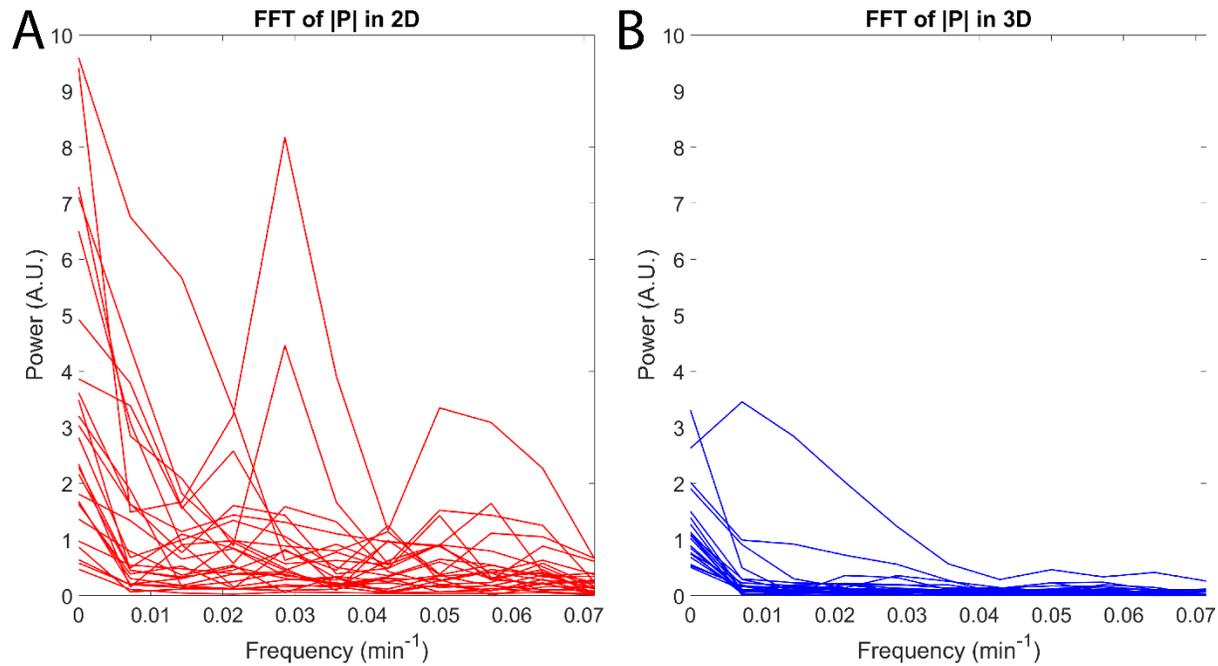

**Figure S6. Periodicity of RhoA polarizations in 2D and 3D. A-B.** Fast Fourier transform of $|P|$ in 2D (left) and 3D (right). In 2D, cells behaved in a heterogenous way as there are subpopulations of cells that are more periodic, but not in 3D. Some cells in 2D showed distinct peak corresponding to period of around 30mins.

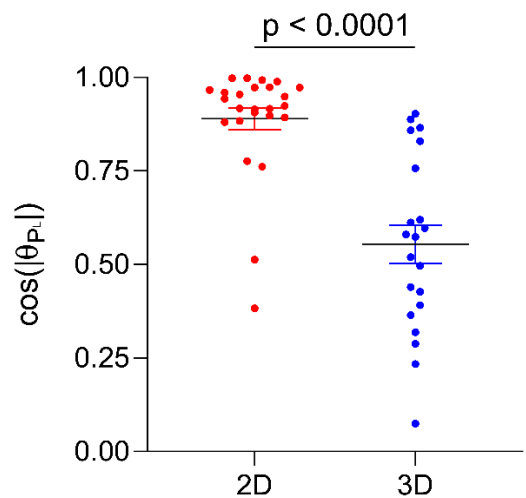

**Figure S7. RhoA Polarization alignment and orientation.** RhoA polarization is more aligned with the long axis of cell in 2D than 3D. The measured absolute angle ($|\theta_{P_L}|$) between polarization vector and the long axis shows that polarization vector and the long axis were significantly more aligned in 2D than in 3D, as indicated by a higher average cosine value. Error bars = SEM.

**Supplementary Video 1.** Cyclic movement of an MDA-MB-231 cell and RhoA polarization during migration on a 2D fibronectin-coated glass substrate. Duration: 70mins; Time interval: 7mins; Image Size: 58.48x58.48$\mu$m.

**Supplementary Video 2.** Non-polarized RhoA activity in an MDA-MD-231 cell migrating in a 3D collagen matrix. Duration: 70mins; Time interval: 7mins; Image Size: 58.48x58.48$\mu$m.